\documentclass[smallcondensed]{svjour3}     

\usepackage{mathptmx}

\usepackage[a4paper,top=2cm,bottom=2cm,left=3cm,right=3cm,marginparwidth=1.75cm]{geometry}

\usepackage{amsmath}
\usepackage{graphicx}
\usepackage{hyperref}

\newcommand*{\affaddr}[1]{#1} 
\newcommand*{\affmark}[1][*]{\textsuperscript{#1}}

\begin{document}

\title{Weak interaction studies at SARAF\thanks{The work presented here is supported by grants from the Pazy Foundation (Israel), Israel Science Foundation (grants no. 139/15 and 1446/16), and the European Research Council (grant no. 714118 TRAPLAB). BO is supported by the Ministry of Science and Technology, under the Eshkol Fellowship.}}

\author{%
		Ben Ohayon\affmark[1]\and
        Joel Chocron\affmark[1]\and
        Tsviki Hirsh\affmark[2]\and
		Ayala Glick-Magid\affmark[1]\and
        Yonatan Mishnayot\affmark[1,2]\and
        Ish Mukul\affmark[3]\and
        Hitesh Rahangdale\affmark[1]\and
        Sergei Vaintraub\affmark[2]\and
        Oded Heber\affmark[4]\and
        Doron Gazit\affmark[1]\and
        Guy Ron\affmark[1]
        }


\institute{
			B. Ohayon \at ben.ohayon@mail.huji.ac.il
            \and
			J. Chocron \at joel.chocron@mail.huji.ac.il 
            \and
            T. Hirsh \at tsviki@soreq.gov.il 
            \and
            A. Glick-Magid \at ayala.glick@mail.huji.ac.il  
            \and
			Y. Mishnayot \at yonatan.mishnayot@mail.huji.ac.il 
            \and
            I. Mukul \at  imukul@triumf.ca 
            \and
            H. Rahangdale \at hitesh.rahangdale@mail.huji.ac.il  
            \and
			S. Vaintraub \at sergeyv@soreq.gov.il 
            \and
            O. Heber \at oded.heber@weizmann.ac.il 
            \and
            D. Gazit \at doron.gazit@mail.huji.ac.il 
            \and
			G. Ron \at gron@racah.phys.huji.ac.il \\\\
            \affaddr{\affmark[1]Racah Institute of Physics, The Hebrew University, Jerusalem 91904, Israel}\\
			\affaddr{\affmark[2]Soreq Nuclear Research Center, Yavne 81800, Israel}\\
			\affaddr{\affmark[3]TRIUMF, 4004 Wesbrook Mall, Vancouver, Canada}\\
            \affaddr{\affmark[4]The Weizmann Institute of Science, Rehovot 76100, Israel}\\
          }

\maketitle
\begin{abstract}
We review the current status of the radioisotopes program at the Soreq Applied Research Accelerator Facility (SARAF), where we utilize an electrostatic-ion-beam trap and a magneto-optical trap for studying the nuclear $\beta$-decay from trapped radioactive atoms and ions. The differential energy spectra of  $\beta$'s and recoil ions emerging from the decay is sensitive to  beyond standard model interactions and is complementary to high energy searches. The completed facility SARAF-II will be one of the world's most powerful deuteron, proton and fast neutron sources, producing light radioactive isotopes in unprecedented amounts, needed for obtaining enough statistics for a high precision measurement.


\end{abstract}

\section{Introduction}
\label{intro}
The differential energy/momentum spectrum of decay products from nuclear $\beta$-decay contains valuable information regarding the nature of interactions governing the decay \cite{2016-FierzOscar}, including possible Beyond Standard Model (BSM) couplings. Here, information is encoded in various coefficients, correlating the spin and or momentum degrees of freedom of the recoiling particles \cite{1957-Jackson}. For an unpolarized sample, two main correlation coefficients govern the decay:
\begin{enumerate}
\item The $\beta$-$\nu$ angular coefficient $a$, which is quadratically sensitive to exotic scalar and tensor interactions for Fermi and Gammow-Teller transitions respectively, as well as scenarios where new couplings are either left- or right-handed  \cite{2018-Review}.
\item The Fierz interference term, denoted $b$, is zero in the SM and linearly sensitive to only  left-handed new physics \cite{2018-Review}.
\end{enumerate}
An effective field theory approach, which provides a theoretical framework in which high-energy searches are compared to precision measurements, shows that as precision in determining these and other correlation coefficients approaches $0.1\%$, decay experiments become complementary and competitive with high-energy searches \cite{2016-EFT,2018-Review}.

Since the (anti)neutrino is virtually impossible to detect, efforts are directed to detection of the $\beta$ energy spectrum, the recoil ion energy spectrum, or some combination of them.

The $\beta$ energy spectrum is relatively simple to measure, since the MeV-scale energies involved make the $\beta$'s less prone to scattering within the sample, and modern detectors measure the energy directly. Assuming that resolution is not an inherent limit, $10^8$ decay events from allowed decay of isotopes with Q-values of order MeV provide sub $0.1\%$ uncertainties in the determination of $b$ \cite{2016-FierzOscar}, and so may place tight constrains on BSM left-handed new physics, provided that uncertainty in various corrections is under control \cite{2018-BetaSpectrum}. On the other hand, the $\beta$ spectrum of allowed transitions is insensitive to $a$, and hence to right-handed new couplings. Interestingly, the spectrum of unique first-forbidden transitions is sensitive to $a$, and so provides a novel scheme for simultaneous constraint of left and right symmetric models \cite{2017-Forbidden}. For the experimental challenges, resulting among others from scattering within the detectors, see for example \cite{2016-WM,2018-Oscar}.

Measurement of the recoil ion energy is more involved, since the roughly 1keV recoils are easily scattered, and so a very dilute sample in ultra high vacuum is required. On the other hand, virtually all recoiling ions can be collected and detected efficiently with a charged-particle detector such as a micro channel plate (MCP). If the daughter nuclei are stable or long-lived, direct high resolution energy detection is not possible by calorimetric methods, and so the spectrum is inferred from kinematic observables such as time of flight (TOF) and hit position distributions, which require either direct or inferred knowledge on the decay position. To address these issues, modern experiments in this field rely on collection of the sample within an atom or ion trap\footnote{A notable exception is the Optical LIthium V-mInus-A experiment (OLIVIA) currently developed at MIT} \cite{2018-Review}. Notwithstanding the experimental challenges, the recoil spectrum is as sensitive to $b$ as $\beta$-spectrum for GT transitions and almost an order of magnitude more sensitive for the case of pure Fermi decays \cite{2016-FierzOscar}. Moreover, it is extremely sensitive to $a$, where our Monte-Carlo analysis shows that approximately $0.1\%$ uncertainty in $a$ is achievable by collecting roughly $10^7$ events, making it an ideal experiment for constraining right-handed new physics. Corrections to allowed transitions, including weak magnetism, radiative and recoil order effects must be taken into account \cite{1974-EMCorrections,1974-Recoil}, as well as decays to forbidden transitions, which are interesting for BSM searches on their own \cite{2017-Forbidden}. 

\begin{figure}
\centering
\includegraphics[width=0.6\textwidth]{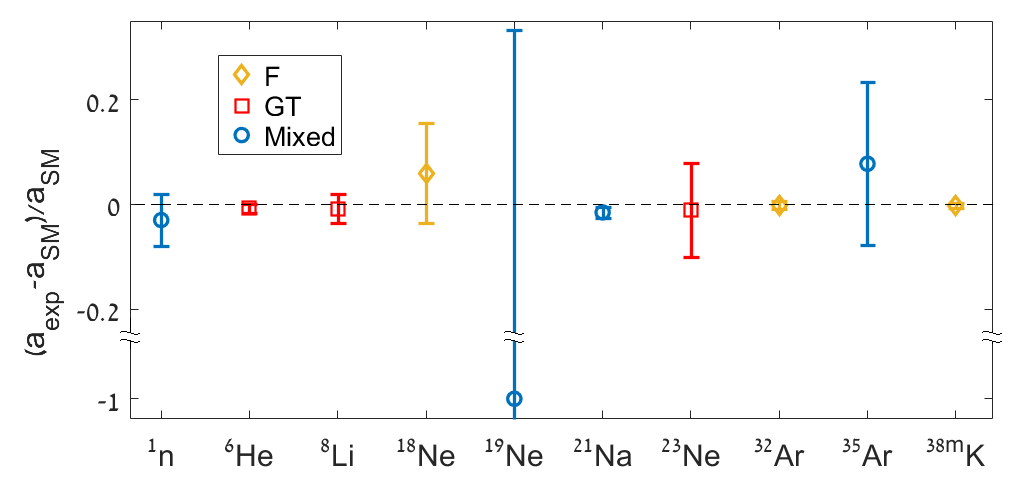}
\caption{\label{fig:a_beta_nu}
Experimentally obtained $\beta$-$\nu$ correlation coefficients compared to Standard-Model value for various isotopes. The most precise value was selected for each isotope. (Taken with permission from \cite{2018-EPJA})}
\end{figure}

The sensitivity to new physics, as well as experimental and theoretical challenges, depends upon the investigated isotope and technique. Figure \ref{fig:a_beta_nu} reviews the most precisely measured values of $a$, for each isotope and the neutron \cite{2018-EPJA}. With the exception of $^6$He \cite{1963-He6OAK,2015-Drake},  all precise determinations stem from experiments involving traps. Inspecting figure \ref{fig:a_beta_nu}, neon isotopes were not measured with high precisions, display diverse modes of decay, and so stand out as promising candidates for trap experiments  \cite{2018-EPJA}.
\section{Studies at SARAF}
\label{sec:1}

\begin{figure}[!ht]
\centering
\includegraphics[width=0.5\textwidth,trim={10 10 5 5},clip]{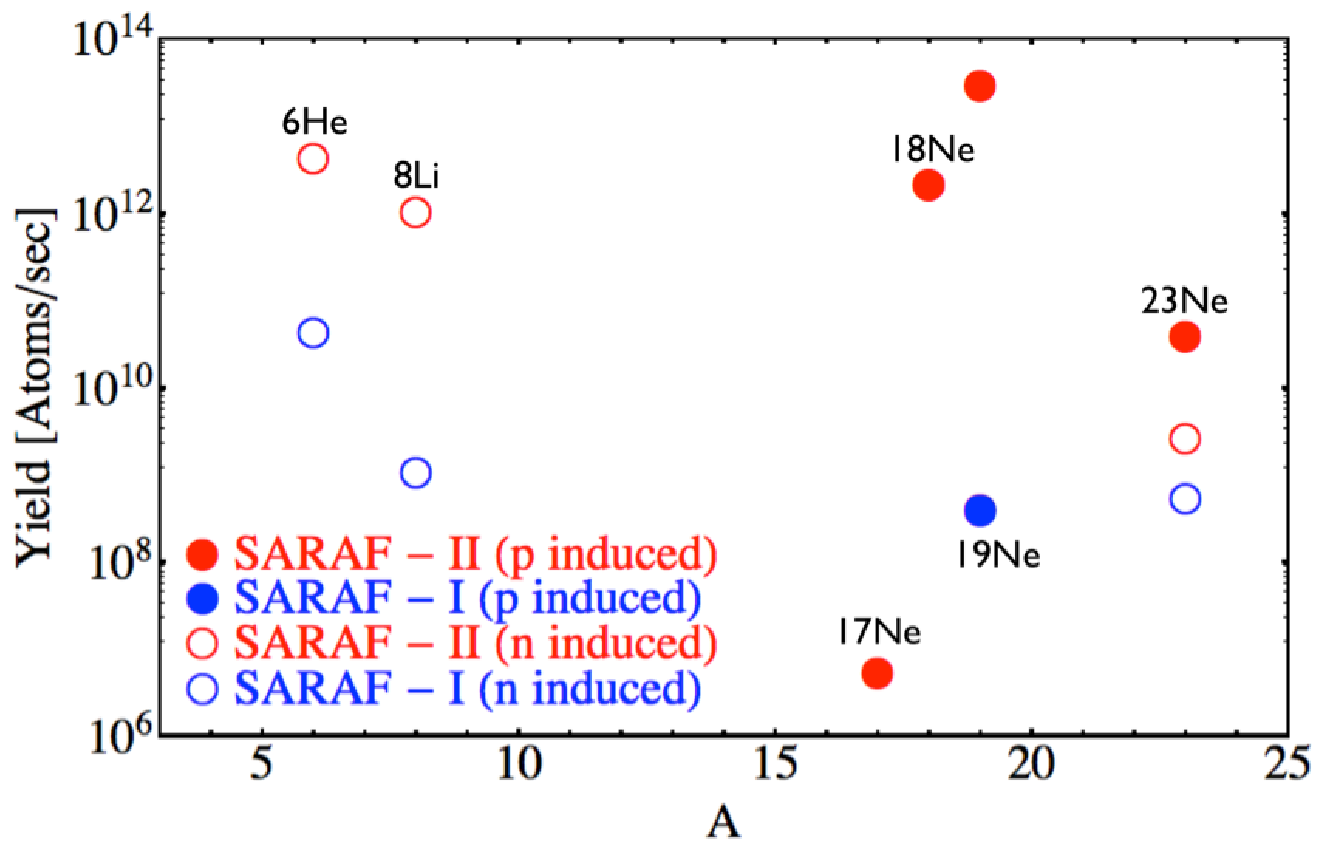}
\caption{\label{fig:Yields}
Expected yield of radioactive atoms of interesting for decay studies at SARAF phase-I and II.}
\end{figure}

The Soreq Applied Research Accelerator Facility (SARAF) is currently being constructed at the Soreq Nuclear Research Center (SNRC). It is designated  as a user facility that will be based on protons and deuterons. Phase-I, which provides $4$ MeV protons/deuterons at maximal continuous current of 2 mA, is currently starting operation. Phase-II, which is under construction, is a medium energy (40 MeV) high CW current (5 mA) accelerator. These specifications, with the proper targets \cite{2018-LiLit}, will make SARAF one of the world's most powerful deuteron, proton and fast neutron sources.  Figure \ref{fig:Yields} provides an overview of expected production rates of light radionuclides of interest to decay studies,  an overview of the facility is found in \cite{2018-EPJA}.

The isotopes of choice must posses an appropriate half-life; long enough so that most reach the trap after production and transport ($ > 100$ ms),  and short enough so that most decay events occur in the trap ($< 100$ s). Since we compare results with those calculated using the standard model, the theory and corrections for these isotopes, as well as other decay properties such as the branching ratios, has to be well-known to the required level of precision ($ < 1$\%). Moreover, the isotopes trapped in the atomic trap should possess a suitable cooling transition where sufficient laser power is available. 

As mentioned earlier, a competitive measurement requires acquisition of a few $10^{7}$ decay events within the trap \cite{2008-ShakeoffVetter}, and so for a reasonable beam time of under 100h, The average event rate should be at least 100/s. For a production rate  of $10^9$ isotopes of choice per second, the overall efficiency of transport, trapping and detection should be at least $10^{-7}$, which is easily obtained in the EIBT setup \cite{2018-Ish}, and is regularly obtained in a state of the art, laser-cooling system for metastables \cite{2017HongCharge}. At the present, a meaningful measurement can be accomplished at SARAF-I with the production capabilities shown in fig. \ref{fig:Yields} for $^{6}$He and $^{23}$Ne. Where the former is to be trapped at an Electrostatic Ion Beam Trap (EIBT) and the latter in a magneto-optical trap (MOT).
Both trap systems have been commissioned and tested with stable isotopes, and moved and reconstructed at SARAF.
\subsection{$^6$He production and trapping in EIBT}
$^{6}$He decays directly to the ground state of $^{6}$Li$^+$ by $\beta^-$ emission, with a Q-value of $3.50521\pm 0.00005$  MeV, and a half life of $806.9\pm0.1$ ms \cite{2012-KnechtHe6}. Our first radioactive atom production experiment for $^{6}$He was performed at the target room of the now-decommissioned Koffler Pelletron Accelerator at the Weizmann Institute of Science and is described in \cite{2018-Ish}. We found a production probability of 1.45$\times$10$^{-4}$ $^{6}$He/neutron, and the total system efficiency (which includes transport and decay during measurement) of nearly 4\%. We are currently commissioning an improved production run at SARAF-I which is expected to yield around $10^{10}$ $^{6}$He/s with subsequent trapping at the EIBT setup \cite{2018-EPJA}.

The EIBT was developed at the Weizmann Institute of Science by the group of Prof. Daniel Zajfman during the 90's \cite{1997-eibt-zafman}.  It consists of two sets of eight electrodes, acting both as electrostatic mirrors and a lens by producing a retarding field, which respectively reflects the beam along its path and focuses it transversely. The innermost electrodes are grounded to create a field-free region ($\sim$24 cm in our case) where the ion beam has a well-defined kinetic energy and direction in space. The entrance mirror is grounded upon injection of the ion bunch. Once the ion bunch fills the entire trap, the potentials on the entrance mirrors are raised quickly ($\sim$50 ns) so that the ion bunch oscillates back and forth between the two mirrors in harmonic motion. Since the trapping is based only on electrostatic repulsion, the EIBT is independent of ion mass and effectively any element can be trapped.  The storage time of such small table-top devices is limited by the residual gas in the chamber \cite{1998-eibt-dahan}. Currently it is about $1$ second, since $^{6}$He half life is around $0.8$ s, more than half of the  trapped ions are expected to decay within the trap. 

Following  $^{6}$He  decay, the electron and recoil ion are measured in coincidence, providing all kinematic information necessary to deduce $a,b$. The position and TOF, and thus the energy of the recoil  $^{6}$Li, are measured by MCP detector with a $6.4$ mm center hole inside the EIBT. The position and energy of the electron are measured by a specially designed large area position sensitive detector, comprising a plastic scintillator and several photomultipliers  \cite{2015-LargeDet}, placed outside of the trap.
The use of such a trap for fundamental interaction studies exhibits potentially very significant advantages over other trapping schemes, since it incorporates an extended field-free region for better kinematic reconstruction, and a large solid angle for the electron and recoiling atom detectors. In addition, since the detector parameters are given, one can optimize the bunch size and positioning to decrease the systematic uncertainties. The overall system efficiency, for ionization in the electron-beam-ion-source and trapping, is currently around $10^{-6}$ \cite{2018-Ish}.

\subsection{ $^{23}$Ne production, branching-ratio and trapping }
$^{23}$Ne decays by $\beta^-$ emission to $^{23}$Na, with a Q-value of $4.3758\pm0.0001$ MeV, and a half life of $37.13\pm0.03$ s \cite{2015-Ne18}. The decay is predominantly to the ground and first excited states and is practically pure GT. The branching ratio (BR) to the aforementioned states was measured by Penning and Schmidt to be $67\pm3\%$, and $32\pm3\%$ \cite{1957-Penning}, and angular correlations were measured most precisely by \cite{1963-carlson}, returning  $a=-0.33\pm0.03$, where most of the uncertainty results from that in the BR. Thus a more accurate measurement of the BR is crucial before any attempt on a more precise determination of $a$ or $b$ is made. During March 2017 we conducted the first production experiment at SARAF phase-I to examine our ability to produce and transport large quantities of $^{23}$Ne via the $^{23}$Na (n,p) reaction. A follow up experiment, dedicated to the measurement of  $^{23}$Ne BR, was recently conducted and is currently being analyzed.

The neon MOT system was commissioned and tested with stable Ne isotopes at the Hebrew University of Jerusalem. Trapping of up to a few $10^6$ atoms at steady-state was demonstrated \cite{2015-ZS}, and coincidence measurements of recoil ions from collisional-ionization mechanisms were performed utilizing the same collection and detection scheme designated for future decay-studies. In order to significantly reduce the background from, and collisions with, ground state radioactive neon, we employ an isotope-selective, forty five degrees deflection stage between the Zeeman slower and science chamber. It has an efficiency of about 30\% and enables fast (<ms) shutting-off of the metastable beam for background determinations.

Whereas most laser-cooling systems require large flux \cite{2011-LuAtta},  or a high density of atoms in the trap necessary to obtain a Bose-Einstein condensate,  our most important quantity is the overall efficiency for cooling and trapping, namely the number of atoms which have decayed within the trap as a fraction of those which have entered the system.  The excitation efficiency to the metastable state for light nobles is $10^{-4}$ at most \cite{2012-BirklReview}, and limits the efficiency of the such system to about $10^{-7} $\cite{2017HongCharge}, and so we are actively pursuing further developments for each part of the system such as the metastable RF source \cite{2015-RFs}, and Zeeman-slower \cite{2013-NewApp,2015-ZS}. Recycling of the gas sample within the system system will be introduced in order to get a  gain of up to $10^3$ in efficiency \cite{2011-LuAtta} for stable isotopes, and our preliminary studies show that a gain of around $50$ is obtainable owing to the relatively long half-life ($37$ s) of $^{23}$Ne.

The MOT lifetime, which, along with the half-life, determines how many of the trapped isotopes decay within the trap and contribute to our statistics. At the low flow associated with the small amounts of radioactive isotopes produced, collisions with residual background gas ($10^{-10}$ Torr), and two-body collisions in the trap, play a minor role, and intrinsic lifetimes become dominant. For stable neon isotopes, this is the effective metastable lifetime of $15$ s \cite{2003-NeLifetime}, which can be extended to up to about $30$ s by increasing the excited fraction to the maximum of one half. In those favorable conditions, slightly less than  $^{23}$Ne atoms will decay within the trap prior to deexciting from the metastable state and being pumped out.

By surrounding the decay volume with a known electric field, the recoil energy spread of an ion translates to a TOF distribution. This TOF is measured either by triggering on the $\beta$'s \cite{2004-Scielzo}, which have high energy and so can not be collected with high geometrical efficiency; or the SO electrons \cite{2008-ShakeoffVetter}, which may have a low probability to be created.  A comparison of MC simulations of ion TOF distribution for different values of $a,b$ results in their determinations and bounds \cite{2008-ShakeoffVetter}.


\subsection{$\beta$ spectrum of $^{16}N$}

As was presented in the introduction, the $\beta$ spectrum of unique first forbidden decays is sensitive to $a$. As a test case, we decided to measure the decay of $^{16}$N, which has a spin parity of $2^-$.   28\% of the time, It decays  to the $0^+$ ground state of $^{16}$O  with Q-value of $10.4$ MeV. Other branches of the decay are pure GT decays to $1^-$ and $3^-$ with Q-values $3.3$ and $4.1$ MeV, respectively. Therefore, in order to measure only the forbidden branch the spectrum one should look for electron energies above $4.1$ MeV. We are planning to measure the spectrum with a high purity germanium detector (HPGe). The detector was calibrated by $^{90}$Sr and $^{207}$Bi sources, and the estimated resolution was found to be roughly $50$ keV.  $^{16}$N will be produced by $^{19}$F(n,a)$^{16}$N reaction on a $100$ micron thin Teflon target, which will be irradiated by a $14$ MeV d-t commercial neutron generator and transported between the generator and HPGe detector within $3$ seconds by a uniquely designed swing device. The experiment is now at final stages of preparation and we expect to start the measurements in about month.

\section{Conclusion}
We presented the light radioisotopes trapping program, at the Saraf applied research accelerator facility, dedicated to the search or constraint of beyond standard model physics in the weak sector. Precise measurements, complementary to high-energy searches, are possible if over $10^7$ decay events are recorded and analyzed, and various corrections are under control. For a reasonable beam time, unprecedented amounts of light radionuclides are required. Over a billion $^{23}$Ne and $^6$He atoms per second are soon to be produced at the low-energy, high-current phase-I of SARAF, and the medium-energy, high-current specifications of the currently constructed phase-II will grant access to others.

\bibliographystyle{abbrv} 
\bibliography{Citations} 

\end{document}